# A Pairwise Comparison Approach to Ranking in Chess Team Championships[1]

**László Csató**

Corvinus University of Budapest, Budapest, Hungary

# A Pairwise Comparison Approach to Ranking in Chess Team Championships

**Abstract**

Chess championships are often organised as a Swiss-system tournament, causing great challenges in ranking the participants due to the different strength of schedules and possible circular triads. The paper suggests that pairwise comparison matrices perform well in similar ranking problems. Some features of the proposed method are illustrated by the results of the 18[th] European Team Chess Championship. The analysis is able to take into account the influence of different opponents and robust with respect to the scaling technique chosen. The method is simple to compute as a solution of a linear equation system.

**Keywords**: Multicriteria decision making, Incomplete pairwise comparison matrix, Chess, Incomplete tournaments, Ranking methods

## Introduction

In the world of sport various evaluation methods have been applied to determine the final ranking of players or teams based on matches between two participants. A kind of substitution for round-robin tournaments is the Swiss-system where all participants face each other for a determined number of games, without a knockout phase. It means that a loss in the first rounds does not make impossible for a contestant to triumph finally. Chess championships are often organised in this way due to the large number of players and tight time limits. It results in two interesting questions: how teams to be paired and how to determine the final ranking.

The paper is confined to the second problem, which simplifies the situation by making possible to use the results of a specific tournament without speculating about a better schedule. Current methods for the final ranking are mainly based on lexicographical orders. They often fail to take into account the results of opponents, addressed by the pairing method and by the secondary components of the lexicographical order, the so-called tie-breaking rules. Chess federations apply many different ranking methods, some of them are strongly criticised (Jeremic and Radojicic 2010)

Pairwise comparison models of Multicriteria Decision Making methodology are able to address similar issues as above. Here the 'alternatives' to be compared are the participants of the tournament, match results will be incorporated into a pairwise comparison matrix, and the final ranking will follow the priority vector derived from the matrix with an estimation method. In the case of Swiss-system tournaments incomplete matrices should be applied.

The detailed analysis will take the results of the 18[th] European Team Chess Championship. This competition took place from November 3[rd] to 11[th], 2011 in Porto Carras, Greece. It is an ideal example for the potential application of incomplete pairwise comparison matrices as participants were interested in defeating their opponents on all boards and distortions due to colour allocation are eliminated.

## Incomplete pairwise comparison matrices

The $n \times n$ real matrix $\mathbf{A} = (a_{ij})$ is a pairwise comparison matrix (Saaty 1980) if it is positive and reciprocal: $a_{ij} > 0$ and $a_{ji} = 1/a_{ji}$ for all $i, j = 1, 2, \ldots, n$. Reciprocity implies that $a_{ii} = 1$ for all $i = 1, 2, \ldots, n$. Matrix element $a_{ij}$ is the numerical answer to the question 'How many times is the $i$th alternative more important than the $j$th?'

The task is to calculate a positive $n$-dimensional weight vector $\mathbf{w} = (w_i)$, where $w_i/w_j$ somehow approximates the pairwise comparison ratio $a_{ij}$. The solution becomes obvious if matrix $\mathbf{A}$ is consistent, namely $a_{ik} = a_{ij} \cdot a_{jk}$ for all $i, j, k = 1, 2, \ldots, n$, when there exists a unique vector $\mathbf{w}$ such that $a_{ij} = w_i/w_j$ for all $i, j = 1, 2, \ldots, n$. Otherwise, $\mathbf{A}$ is inconsistent. In sport circular triads ($A$ beats $B$, $B$ beats $C$, and $C$ beats $A$) are often present, implying inconsistency.

In this case the real ranking of players can be estimated at most. One method for this purpose is the Logarithmic Least Squares Method (*LLSM*, Crawford and Williams 1985), which minimizes the function

$$\sum_{i=1}^{n} \sum_{j=1}^{n} \left[ \log a_{ij} - \log\left(\frac{w_i}{w_j}\right) \right]^2 \to \min$$

subject to $\sum_{i=1}^{n} w_i = 1$, $\mathbf{w} \in \mathbb{R}_+^n$. Its main advantage is simple computation of the optimal weights as the geometric means of row elements.

Incomplete pairwise comparison matrices are defined if there are some missing elements outside the diagonal (Harker 1987). The application of *LLSM* to the incomplete case is achieved by omitting the unknown elements of $\mathbf{A}$ from the objective function (Bozóki et al. 2010). The solution is unique if and only if all alternatives are compared directly or indirectly, that is, matrix $\mathbf{A}$ is irreducible. Optimal weights are the solution of a linear equation system.

# Application to the European Team Chess Championship 2011

Our incomplete pairwise comparison matrix is given by the results of the championship described in the introduction (Chess Results 2011). The 38 participating teams played 9 matches respectively, so 171 elements are known compared to the total 703 in the upper triangle of a 38×38 matrix (24.3%).

A match between two teams consists of games on 4 boards with 3 possible results (white win or loss, draw). The winner of a game gets 1 game point, the loser 0, while a draw means 0.5 game points for both players. The sum of game points (called board points), ranges from 0 to 4, by 0.5. The team which scores more board points on a match receives 2 match points, the opponent scores no match points. In case of equal board points, each team receives 1 match point. The distribution of results is presented in Figure 1.

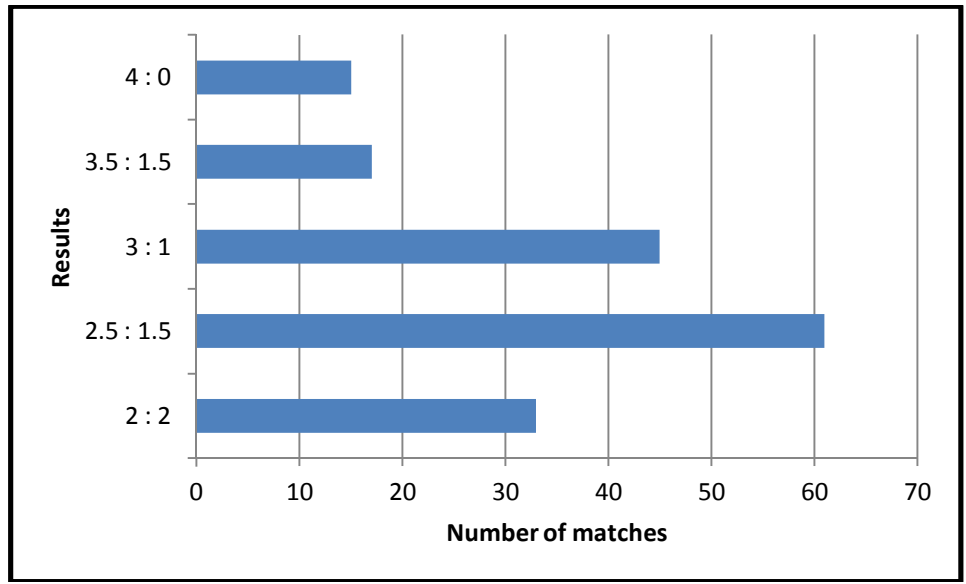

**Figure 1: Distribution of match results**

The official ranking will be made according to match points (TB1). Tie-breaks shall be determined by application of 4 tie-breaking procedures in sequence, proceeding from TB2 to TB5 to the extent required (European Chess Union Tournament Rules 2012). As in 9 matches at most 18 match points could be scored, TB rules certainly should be applied, thus teams are strongly interested in increasing their game points as TB2 is the number of board points. Consequently, it is justified to give higher weight to wins scored with more game points.

Match results should be transformed into values (ratios) to fit into a pairwise comparison matrix. Draws (2:2) are obviously converted to 1, for the other results 4 different rules were applied, presented in Table 1 (since reciprocity it is enough to see the results from the winner's point of view).

| Result | PC1 | PC2 | PC3 | PC4 |
|---|---|---|---|---|
| 2.5 : 1.5 | 2 | 3 | 3 | 3 |
| 3 : 1 | 3 | 5 | 4 | 3 |
| 3.5 : 1.5 | 4 | 7 | 5 | 3 |
| 4 : 0 | 5 | 9 | 6 | 3 |

**Table 1: Transformation of match results into pairwise comparison ratios**

The number of game points counts most in PC2, while PC4 applies a simple binary coding for victories. PC1 is the baseline scenario, PC3 represents a transition between PC2 and PC4.

The irreducibility of the incomplete pairwise comparison matrix is intuitively provided due to the pairing method of the tournament (European Chess Union Tournament Rules 2012), and our check back this idea.

## Analysis of the rankings

The *LLSM* method provides an optimal weight vector for the teams used as a basis of our rankings. With a slight abuse of notation, PC1 / PC2 / PC3 / PC4 will correspond to the ranking derived from the analogous matrices. Start serves as a reference, it is the ranking of teams before the tournament determined by the average of the 4 highest FIDE rating of their players, reflecting former individual performance. Final is the official final result of the championship (Chess Results 2011).

One of the most known index for comparing rankings is Spearman's rank correlation coefficient (Spearman 1904). It is an element of the $[-1, +1]$ interval, the limits are reached when the two rankings are exactly the same $(+1)$ or entirely opposite $(-1)$.

|  | Start | Final | PC1 |
|---|---|---|---|
| Start | 1 | 0.8718 | 0.9223 |
| Final | 0.8718 | 1 | 0.9431 |
| PC1 | 0.9223 | 0.9431 | 1 |

**Table 2: Pairwise rank correlations between some rankings**

Rank correlations between our 4 rankings are above 0.98, PC4 is a bit outsider as it uses a simple binary coding for wins. It implies that pairwise comparison methods have common roots and the proposed rankings are robust regarding the transformation of match results. Seemingly high values in Table 2 could be deceptive because the ability of team's players (Start) significantly determines the final outcome. It means the Final and suggested rankings are relatively different, which means we should look for other indicators to judge the validity of the calculated rankings.

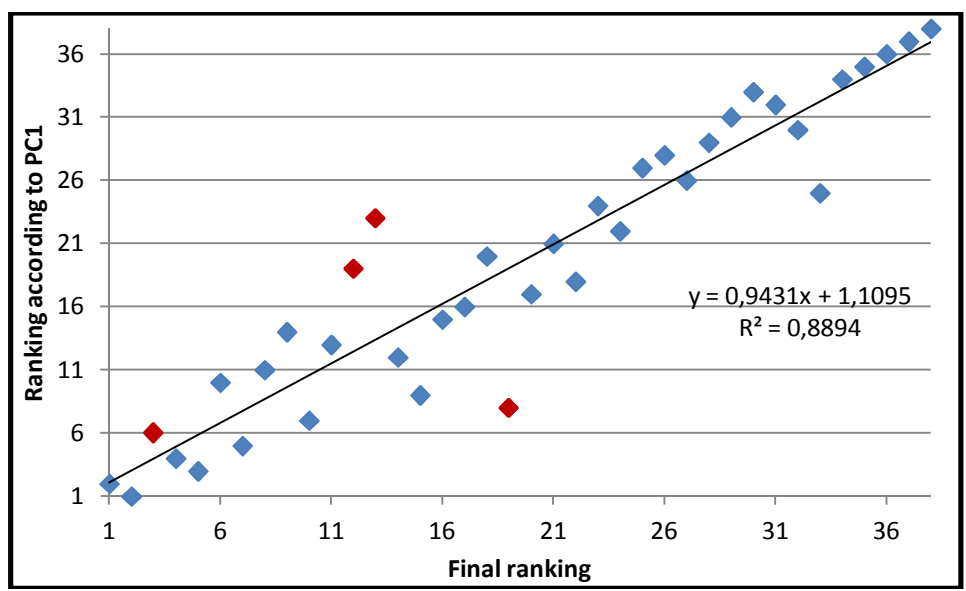

**Figure 2: Relation of Final and PC1 rankings**

The position of teams in Final and PC1 rankings are drawn on Figure 2. The main discrepancies are for the following countries (marked by red squares):
- France, which is 19$^{th}$ in Final, but at least 10$^{th}$ in our rankings;
- Hungary, which is 3$^{rd}$ in Final, but 6$^{th}$ in our rankings;
- Serbia, which is the 12$^{th}$ in Final, but at most 19$^{th}$ in PC1-PC4;
- Georgia, which is 13$^{th}$ in Final, but at most 23$^{rd}$ in PC1-PC4.

These differences arise because the results of opponents are not reflected by the official method. France had an unfavourable schedule as it was among the best in the first rounds, meaning to play against strong teams, contributing to the 'exhaustion' of players resulting in a sharp decline later. The other three countries proceeded on the 'outer circle': they were able to exploit the opportunity given by weaker initial performance through defeating the more favourable opponents. It is shown in their TB3 points (the sum of the board points of all opponents), placing France to 7[th], but Hungary to 16[th], Serbia to 29[th] and Georgia to 33[rd].

Our rankings are relatively robust among the best: the first six positions are occupied by the first 7 teams in Final. Only Armenia and Bulgaria swap places in the 4[th] and 5[th] positions. It strengthens the view that the pairwise comparison approach is independent from the transformation of results. However, the winner (Azerbaijan) in PC1-PC4 is different from the winner according to the official ranking (Germany). We were able to identify three possible factors, which contribute to Azerbaijan's outgunning of Germany's higher match points:

1) Azerbaijan has won its matches with a greater margin; its number of board points (TB2) is 23 versus 22.5 for Germany. It is reinforced by the observation that the ratio of their optimal weights is the smallest in PC4.
2) There is a circular triad among Azerbaijan, Bulgaria and Germany with Azerbaijan-Bulgaria 3.5:0.5, Bulgaria-Germany 3:1 and Germany-Azerbaijan 2.5:1.5. It is more favourable for Azerbaijan because of the game points of wins.
3) Germany has a 2:2 result against Israel, a weaker team by our method than Azerbaijan's opponents for its two draws, Spain and France.

## Conclusion

This paper presents an alternative method to determine the final ranking of a Swiss-system tournament based on incomplete pairwise comparison matrices. Rankings according to the *LLSM* method are robust with respect to the arbitrary scales for the transformation of match results. Our proposal is able to take into account the results of opponents and reflects the varying margin of wins. It can provide a unique opportunity to further analyze chess tournaments and improve the final ranking in order to receive intuitively better results.

## References

Bozóki, S., Fülöp, J. & Rónyai, L. 2010. On optimal completion of incomplete pairwise comparison matrices. *Mathematical and Computer Modelling* 52(1-2): 318-333

Chess results 2011. Retrieved from
http://www.chess-results.com/tnr57856.aspx?lan=1 on 11[th] March, 2012.

Crawford, G. and Williams, C. 1985. A note on the analysis of subjective judgment matrices. *Journal of Mathematical Psychology* 29: 387-405

European Chess Union Tournament Rules 2012. Retrieved from http://europechess.net/index.php?option=com_content&view=article&id=9&Itemid=15 on 11th March, 2012.

Harker, P.T. 1987. Incomplete pairwise comparisons in the analytic hierarchy process. *Mathematical Modelling* 9(11): 837-848

Jeremic, V. M. and Radojicic, Z. 2010. A new approach in the evaluation of team chess championships rankings. *Journal of Quantitative Analysis in Sports* 6(3): article 7

Saaty, Thomas L. 1980. *The analytic hierarchy process*. New-York, NY: McGraw-Hill

Spearman, C. 1904. The proof and measurement of association between two things. *American Journal of Psychology* 15: 72-101